# On the evolution of character usage of classical Chinese poetry

Lili Yu[1] and Liang Liu[2*]

[1]Department of Biostatistics, College of Public Health, Georgia Southern University, Statesboro, GA 30303

[2]Department of Statistics and Institute of Bioinformatics, University of Georgia, Athens, GA 30606

*Corresponding author:

Liang Liu (lliu@uga.edu)

Department of Statistics

Institute of Bioinformatics

University of Georgia

101 Cedar Street

Athens, GA 30602, USA

Tel: 706-542-3309

**ABSTRACT**

The hierarchy of classical Chinese poetry has been broadly acknowledged by a number of studies in Chinese literature. However, quantitative investigations about the evolutionary linkages of classical Chinese poetry are limited. The primary goal of this study is to provide quantitative evidence of the evolutionary linkages, with emphasis on character usage, among different period genres of classical Chinese poetry. Specifically, various statistical analyses are performed to find and compare the patterns of character usage in the poems of nine period genres, including *shi jing*, *chu ci*, *Han shi* , *Jin shi*, *Tang shi*, *Song shi*, *Yuan shi*, *Ming shi*, and *Qing shi*. The result of analysis indicates that each of nine period genres has unique patterns of character usage, with some Chinese characters that are preferably used in the poems of a particular period genre. The analysis on the general pattern of character preference implies a decreasing trend in the use of Chinese characters that rarely occur in modern Chinese literature along the timeline of dynastic types of classical Chinese poetry. The phylogenetic analysis based on the distance matrix suggests that the evolutionary linkages of different types of classical Chinese poetry are congruent with their chronological order, suggesting that character frequencies contain phylogenetic information that is useful for inferring evolutionary linkages among various types of classical Chinese poetry. The estimated phylogenetic tree identifies four groups (*shi jing*, *chu ci*), (*Han shi*, *Jin shi*), (*Tang shi*, *Song shi*, *Yuan shi*), and (*Ming shi*, *Qing shi*). The statistical analyses conducted in this study can be generalized to analyze the data sets of general Chinese literature. Such analyses can provide quantitative insights about the evolutionary linkages of general Chinese literature.

## 1. INTRODUCTION

Quantitative measurements have been commonly used in linguistic studies for understanding various language phenomena and language structures (Liu and Huang 2012). The analysis of affinity among Chinese dialects adopted correlation coefficients to quantify dialect similarity (Cheng 1991). Peng et al. (2008) utilized average distances, clustering coefficients, and the degree distribution to quantitatively compare the networks of Chinese syllables and characters. Moreover, it has become common practice in quantitative linguistics to test language laws on empirical data using statistical methods. Increasing involvement of statistics and mathematics in linguistic studies has significantly changed the way of conducting scientific research for understanding important aspects of language. The origins and development of human language are perhaps the most fundamental questions in linguistics. A closely related but much more complicated question is how human language evolves over time. Various hypotheses regarding the origins and evolution of human language have been proposed and described in the context of probabilistic and biological models (Freedman and Wang 1996; Wang 2013). Since August Schleicher first introduced the representation of language families as an evolutionary tree, phylogenetic trees have been fundamental tools for understanding the history of language in the context of human evolution (Wang 1998; Wang 1998). Neighbor-joining (Saitou and Nei 1987) is one of the most commonly used methods for building phylogenetic trees in evolutionary linguistics. In addition, the efficiency of the algorithm searching for the best tree was greatly improved by matrix decomposition of phylogenetic trees (Qiao and Wang 1998).

Quantitative analysis of poetic texts is related but distinct from the quantitative studies in evolutionary linguistics. The quantitative analysis of poetic texts particularly aims to understand textual patterns of poetry of various forms or genres. The statistical analysis of classical Chinese poetry has shed light on the political, social, and economical status of Chinese historical periods. Particularly, using statistical tools for textual analysis, a study on Tang poems has revealed the important role of geography, history, and architecture in the development of poems of Tang dynasty (Lee and Wong 2012). The textual analysis of Chinese poems collected from the 1920s through the 1990s suggests that the classical character of Chinese poetry declined across the 20th century (Voigt and Jurafsky 2013). Quantitative analysis of poetic texts has gained popularity in comparing similarities and distinctions among various styles of poetry (Popescu et al. 2011; Popescu et al. 2015). A number of statistical tools, including discriminant (Andreev 2003) and network analyses (Roxas-Villanueva et al. 2012), have been employed to characterize poetic style (Popescu et al. 2012; Popescu et al. 2013) or to estimate similarity between poetic texts and their translations (Irma 2007). This article employs statistical tools for textual analysis in the context of phylogenetic trees to understand the unique patterns and evolution of character usage across classical Chinese poetry of nine dynastic genres.

Classical Chinese poetry is one of the most important components of classical Chinese literature. Unlike other forms of classical Chinese literature, classical Chinese poetry was generally developed to be chanted, especially folk poetry, which was orally composed and spread among illiterate folk (Hinton 2008). The earliest Chinese poem *tan ge* (谈歌) was first cited in an archer's response to an inquiry of the king *Gou Jian* of Yue about the secret of accurate bow shooting. Although this short poem was recorded in a history book *wu yue chun qiu*《吴越春秋》dated back to 80 C.E., historians speculated that the poem was actually passed down orally from Chinese primitive society (2600 B.C.E – 2100 B.C.E) and was documented in books much later by descendants. In general, Chinese poetry consists of two types of poetry, namely, classical

Chinese poetry and modern Chinese poetry (Yip 1997). Classical Chinese poetry is characterized as written in traditional Chinese with certain traditional modes associated with particular historical periods (Hinton 2008). The tradition of classical Chinese poetry began at least as early as the publication of *shi jing* (i.e., the Book of Songs), a collection of 305 poems from over two millennia ago (Watson 1984). Classical Chinese poetry continued to grow up until May fourth 1919 movement (Grieder 1980), which is commonly considered as stimulus of emergence of modern Chinese poetry (Yeh 1991).

Classical Chinese poetry has formed its unique style of rhythm and character usage (Zhong 2010). The earliest anthology of classical Chinese poetry is *shi jing*, in which poems or songs are predominantly composed of four-character lines developed between Western Zhou Dynasty and the Spring and Autumn period (Dobson 1964). There are three chapters in *shi jing*, namely, folkway *feng*, elegance *ya*, and praise *song*. The chapter of folkway primarily features folk songs, while the chapter of elegance includes songs from high-class officials and nobility. The songs in the chapter of praise are predominantly used to sing with dance in the monarch ritual of worship for ancestors. Another early anthology of Chinese poetry *chu ci*, known as the songs of the South, consists of poems particularly associated with the state of Chu in southern china. Most poems in *chu ci* are attributed to Qu Yuan and Song Yu. In contrast to *shi jing*, *chu ci* is typified by irregular line lengths (Hawkes 2011). The poems in *shi jing* and *chu ci* reflect, to some extent, the social and political status of pre-Qin period. A new form of classical Chinese poetry, known as *yue fu* style, was developed during the Han Dynasty (Birrell 1993). In contrast to *shi jing* and *chu ci*, *yue fu* poems are composed of five-character lines or seven-character lines (Liu 1966). The *yue fu* style poetry developed during the Han and Jian'an period (the end of the Han Dynasty and the beginning of the Six Dynasties era) later became known as *gu ti shi* (i.e., ancient style poetry) (Watson 1971) in order to distinguish from the poetry developed during the Tang dynasty up to Qing dynasty, which are known as *jin ti shi* (i.e., new style poetry) (Yip 1976). Unlike its predecessor (i.e., ancient style poetry), new style poetry is being composed of five-character or seven-character lines with strict rules for the number of lines, rhyme, and a certain level of mandatory parallelism (Graham 1977). A study by Mair and Mei (1991) has shown that tonal patterns of new style poetry rooted in Sanskrit prosody and poetics (Deo 2007).

Classical Chinese poetry continued to evolve in the context of social, political, and cultural changes over the past three thousand years (Olga 2013). The hierarchical structure of classical Chinese poetry has been broadly acknowledged by a number of studies in Chinese literature (Yip 1976). Two ancient anthologies, *shi jing* and *chu ci*, are on the top of the hierarchy, followed by *yue fu* style poetry (Han dynasty) and new style poetry (Tang dynasty - Qing dynasty), which represents the history of classical Chinese poetry along the lines of dynastic change. However, quantitative investigations about the evolutionary linkages of classical Chinese poetry are exceedingly limited. The primary goal of this article is to provide quantitative evidence of the evolutionary linkages, with emphasis on character usage, among different period genres (Table 2) of classical Chinese poetry. Specifically, statistical analyses are performed to illustrate the unique patterns of character usage across classical Chinese poetry of nine dynastic genres. In addition, phylogenetic trees are used as primary tools to investigate the evolutionary linkages of poetic styles in character usage. Specifically, the similarity of character usage is measured by the average distance between the frequencies of Chinese characters in the classical Chinese poems of nine period genres. A neighbor-joining tree is then built from the matrix of similarity scores to estimate the evolutionary linkages of character usage among nine period genres. Classical Chinese poetry of nine period genres are classified into groups in the estimated phylogenetic tree. The phylogenetic tree of classical Chinese poetry is compared with the chronical order of the dynastic periods from which the poetry was developed.

## 2. Methods and Results

The data set of classical Chinese poems was collected from nine period genres in chronological order, including *shi jing*, *chu ci*, *Han shi*, *Jin shi*, *tang shi*, *Song ci*, *Yuan shi*, *Ming shi*, and *Qing shi* (Table 2). Specifically, the data set consists of 305 poems from *shi jing* and 15 poems from *chu ci*, 675 *Han shi* from the Book of Han and the Records of the Grand Historian, and 1821 *Jin shi* were selected from the Book of Jin. The data set also contains 5000 *Tang shi* sampled from the book *Quan Tang Shi* (Yin et al. 1706), 5000 *Song shi* sampled from *Quan Song Shi database* (Li 2005), 8651 *Yuan shi* from the book *Yuan Shi Bie Cai Ji* (Zhang 2012) and *Classical Chinese Poetry database* (Fang 2014), 14602 *Ming shi* from the book *Lie Chao Shi Ji* (Qian 2007), and 27801 *Qing shi* from the book *Qing Shi Hui* (Xu 1929). The whole corpus contains a total of 4,638,370 Chinese characters collected from *shi jing* (30384 characters), *chu ci* (26529 characters), *Han shi* (27668 characters), *Jin shi* (77715 characters), *Tang shi* (306848 characters), *Song shi* (260622 characters), *Yuan shi* (575968 characters), *Ming shi* (1277619), and *Qing shi* (2055017 characters) (Table 2). The statistical analyses include: (1) the analysis of the distributions of character usage in the poems of nine period genres; (2) the significant analysis of character usage in the poems of nine period genres; (3) the similarity analysis on the evolutionary linkages of character usage among the poems of nine period genres.

In the analysis of the distribution of character usage, I calculated the frequency of each Chinese character occurring in the poems of nine period genres, respectively (Figure 1). The Chinese characters were ordered in accordance with their frequencies from highest to lowest. Moreover, twenty most frequently used Chinese characters (MFUCC) were selected for downstream analyses (Figure 1). The analysis consists of two types of comparisons among the poems of nine period genres. The comparison among the distributions of character usage suggests that the character frequencies of *shi jing* and *chu ci* are significantly different from those of other period genres (Figure 1). The distributions of character frequencies in *Han shi* down to *Qing shi* are relatively flat, indicating that most characters are used with similar frequencies in *Han shi*, *Jin shi*, *tang shi*, *Song shi*, *Yuan shi*, *Ming shi*, and *Qing shi* (Figure 1). In contrast, some characters in *shi jing* and *chu ci* are more frequently used than other characters. This pattern is even more obvious in *chu ci*, where the character *xi* "兮" dominates the distribution of character usage (Figure 1). The high frequencies of *zhi* "之"， *er* "而" in *shi jing* and *chu ci* are due to the fact that *shi jing* and *chu ci* were written in classic Chinese, which is characterized by the high usage of *zhi* and *er*. In addition, the frequent use of *xi* "兮" is the unique feature of *chu ci* style, which explains the high frequencies of *xi* observed in the book of *chu ci*. This result indicates that the Chinese language has profound influence on classical Chinese poetry, which will be further discussed in other analyses. Many MFUCC of *shi jing* and *chu ci* are syncategorematic characters, while most MFUCC of other period genres are notional characters (i.e. noun). In addition, *chu ci* employs *yu* "余" to denote the first person, but other period genres use the character *wo* "我" as the first person. The character "兮" is included in the MFUCC of *shi jing*, *chu ci*, *Han shi*, but not in *Jin shi*, *Tang shi*, *Song shi*, *Yuan shi*, *Ming shi*, and *Qing shi* (Figure 1). The most frequently used characters (one, wind, cloud, sky) are shared by *Jin shi*, *Tang shi*, *Song shi*, *Yuan shi*, *Ming shi*, and *Qing shi* (Figure 1), which is an interesting pattern of post-Han *shi* in their preference towards particular characters.

In the significant analysis of character usage among the poems of nine period genres, I calculated the frequency of each character in the poems of a particular period genre. I further calculated the frequency of the same character in the poems combined across all other period genres. The Binomial hypothesis test (Howell 2007) was adopted to evaluate if the frequency of a character in the poems of a particular period genre was significantly higher than other period genres. Since the test involved multiple comparisons, Bonferoni correction (Dunnett 1955) was carried out to adjust the α level of the Binomial test, leading to the adjusted $\alpha_{adj}$ = 1e-30. Thus, if P-value $\leq\alpha_{adj}$, it was concluded that the frequency of the character in the poems of a particular period genre was significantly higher than other period genres, and this character was identified as a "characteristic character". The result of the analysis shows that the number of characteristic characters in *Jin shi*, *Tang shi*, and *Ming shi* are significantly (P-value < 0.05) less than the numbers of characteristic characters in other period genres. The number of characteristic characters indicates the degree of uniqueness of a particular style of poetry in character usage. The result suggests that *shi jing*, *chu ci*, *Han shi*, *Song shi*, *Yuan shi*, and *Qing shi* have unique styles of character usage, because they have a high number of characteristic characters. In contrast, *Jin shi*, *Tang shi*, and *Ming shi* do not appear to have unique styles of character usage. The spectrum of the degree of uniqueness of nine period genres in Figure 2 indicates four groups, (*shi jing*, *chu ci*), (*Han shi*), (*Song shi*, *Yuan shi*), and (*Qing shi*) that have developed unique styles of character usage. For example, *Han shi* represents ancient style poetry, and (Song shi, *Yuan shi*) represents new style poetry, while *Jin shi* and *Tang shi* are in transition from ancient style poetry to new style poetry. Thus, the spectrum describes a continuous process of evolving from one style to another style during the history of Chinese classical poetry (Figure 2). In addition, the spectrum in Figure 2 indicates that the poems of nine period genres may be divided into four groups, (*shi jing*, *chu ci*), (*Han shi*), (*Song shi*, *Yuan shi*), and (*Qing shi*). *Jin shi* and *Tang shi* are the boundary points between ancient style poetry represented by (*Han shi*) and new style poetry represented by (*Song shi*, *Yuan shi*). Further phylogenetic analyses (see Figure 3) suggest that *Tang shi* is grouped with (Song shi, *Yuan shi*), while *Jin shi* is grouped with *Han shi*. Moreover, *Qing shi* appears to form a separate group and *Ming shi* is in the period of transition from group (*Song shi*, *Yuan shi*) to group (*Qing shi*). The phylogenetic analysis suggests that *Ming shi* is the sister lineage of *Qing shi* and together they form a unique group (see Figure 3). The spectrum in Figure 2 is largely consistent with the evolutionary changes of the Chinese language, which has shown from frequency analysis high impact on the character usage of classical Chinese poetry. The Chinese language evolved from classic Chinese to modern Chinese at the end of Qing dynasty. Moreover, Old Chinese, the early form of classic Chinese, evolved to Middle Chinese at the end of Han dynasty. *Shi jing* and *chu ci* were written in Old Chinese dated from the Western Zhou period to the end of Han dynasty. Thus, *Shi jing* and *chu ci* have a relatively high number of characteristic characters that are not shared by other classical Chinese poetry (Figure 2). Additionally, Qing dynasty is the transition period of the Chinese language from classic Chinese to modern Chinese and the data collected in this paper do not include the Chinese poetry after Qing dynasty. Thus, *Qing shi* shows a high number of characteristic characters that are not shared by other classical Chinese poetry (Figure 2).

The first ten characteristic characters（“埶”, “靦”, “駜”, “僣”, “鳲”, “萑”, “涖”, “襛”, “觩”, “鋈”）of *shi jing* are Chinese characters with particular meanings that rarely occur in the poems of post-Qin period. Thus, the poems of *shi jing* have a unique style of character usage, which can be used as evidence to distinguish from the poems of other period genres. The first ten characteristic characters（“後”, “悽”, “兮”, “嗽”, “些”, “徃”, “塊”, “嵾”, “恒”, “硃”）of *chu ci* includes the character *xi* “兮”, which is consistent with the conclusion of the previous analysis that frequent usage of character *xi* is an unique feature of *chu ci*. Another characteristic character of *chu ci* is *qi* “悽” (miserable), which reflects the miserable mood throughout the poems of *chu ci*. The characteristic characters of *Han shi*, *Jin shi*, and *Yuan shi* include (“僠”, “尿”, “巛”, “睮”, “矸”, “侙”, “唌”, “[illegible]”, “噁”, “塉”), (“轜”, “隟”, “咧”, “腩”, “盻”, “刲”, “哅”, “堦”, “岥”, “崐”), and（“楱”，“檥”，“琱”，“鞹”，“騣”，“髃”）, respectively. It appears that most characteristic characters of *Han shi*, *Jin shi*, and *Yuan shi* are rarely used in modern Chinese literature. In contrast, most of the characteristic characters of *Ming shi* (“沉”, “灯”, “着”, “闲”, “剪”, “胡”, “宫”, “只”, “托”, “醉”）and *Qing shi* (“镫”, “字”, “号”, “沈”, “一”, “間”, “古”, “诗”, “讬”, “奇”) are frequently-used characters in modern Chinese literature. This result shows the tendency of using less rarely used Chinese characters (in modern Chinese literature) along the lines of dynastic changes.

Finally, in the similarity analysis, the pairwise similarity scores were calculated among the poems of nine period genres. The dissimilarity scores were calculated by a distance function defined as the summation of differences between the frequencies of Chinese characters used in the poems of two period genres. Let $x_{1k}$ and $x_{2k}$ be the frequency of the $k^{th}$ character used in the poems of period 1 and period 2, respectively. The distance $D$ between period 1 and period 2 was equal to $D=\sum_k |x_{1k}-x_{2k}|$. Since there were nine periods, the pairwise distances form a nine-by-nine distance matrix. The phylogenetic tree was reconstructed from the distance matrix using the Neighbor-joining method (Saitou and Nei 1987). Moreover, the support values of the estimated phylogenetic tree were calculated using bootstrap techniques (Efron 1971). Specifically, a Neighbor-joining tree was reconstructed for each of 100 bootstrap samples randomly generated with replacement from the original data set. The support value of a branch of the estimated phylogenetic tree was equal to the number of the bootstrap trees containing the branch.

The branch lengths of the estimated phylogenetic tree represent the distances among nine period genres (Figure 3). If two period genres are located nearby in the tree (for example, *Ming shi* and *Qing shi* in Figure 3), it indicates high similarity between the two period genres. Thus, the phylogenetic tree describes the evolutionary relationships among nine period genres for classical Chinese poetry. In figure 3, the phylogenetic tree built from the distance matrix is highly supported with all bootstrap support values = 100. The topology of the tree is consistent with the dynastic order of nine period genres, except that *Yuan shi* and *Song shi* form a clade, which does not reflect or conflict with the chronological order of Yuan and Song dynasty. There are three major groups (*shi jing, chu ci*), (*Han shi*, *Jin shi*), and (*Tang shi*, *Song shi*, *Yuan shi*, *Qing shi*, *Ming shi*) in the phylogenetic tree estimated from the distance matrix (Figure 3). Two major groups, (*Han shi*, *Jin shi*) and (*Tang shi*, *Song shi*, *Yuan shi*, *Qing shi*, *Ming shi*), are consistent with the current classification of classical Chinese poetry (i.e., ancient style poetry and new style poetry) (Olga 2013). However, the phylogenetic tree and the analysis of character usage in Figure 2 suggest that new style poetry may further divide into two separate groups (*Tang shi*, *Song shi*, *Yuan shi*) and (*Qing shi*, *Ming shi*).

The phylogenetic analysis indicates that the frequencies of Chinese characters vary among three major groups (*shi jing, chu ci*), (*Han shi*, *Jin shi*), and (*Tang shi*, *Song shi*, *Yuan shi*, *Qing shi*, *Ming shi*). A number of factors may contribute to the significant variation of the usage of Chinese characters among classical Chinese poetry. Previous studies in evolutionary linguistics indicate that Chinese languages have evolved for over three thousands of years. Old Chinese is the earliest form of the Chinese language used from the Western Zhou period to the end of Han dynasty. Old Chinese differs from later varieties of Chinese in pronunciation, grammar, and usage of Chinese characters. Two earliest Chinese poetry, *shi jing* and *chu ci*, are written in Classic Chinese (the written form of Old Chinese), which differentiates the usage of Chinese characters in *shi jing* and *chu ci* from those in other classical Chinese poetry. Middle Chinese is the Chinese language used from Northern and Southern dynasties to the Song dynasty. Middle Chinese has unique characteristics that may contribute to significant dissimilarities in usage of characters between two groups (*Tang shi*, *Song shi*, *Yuan shi*) and (*Qing shi*, *Ming shi*). Since *Han shi* and *Jin shi* were written during the transition period (Han and Jin dynasties) from Old Chinese to Middle Chinese, the most frequently used characters in *Han shi* and *Jin shi* appear to be the mixture of (*shi jing*, *chu ci*) and (*Tang shi*, *Song shi*, *Yuan shi*) (Figure 1). The most frequently used characters in classical Chinese poetry are consistent with the evolutionary pattern of the Chinese language, evolving from classic Chinese to modern Chinese (Figure 1). This result suggests, with no surprise, that the evolutionary pattern of the Chinese language be inherited by classical Chinese poetry. In other words, the evolutionary changes of the Chinese language are the major contributors to the classification of classical Chinese poetry into three major groups (*shi jing*, *chu ci*), (*Han shi*, *Jin shi*), and (*Tang shi*, *Song shi*, *Yuan shi*, *Qing shi*, *Ming shi*). As *Ming shi* and *Qing shi* were written during the transition period of the Chinese language from classic Chinese to modern Chinese, the frequencies of Chinese characters used in *Ming shi* and *Qing shi* appear to be more similar to Modern Chinese. It leads to the separation of (*Qing shi*, *Ming shi*) from group (*Tang shi*, *Song shi*, *Yuan shi*, *Qing shi*, *Ming shi*). The traditional classification of classical Chinese poetry is based on the number of lines, rhyme, and a certain level of mandatory parallelism, which results in two categories, namely, the old style poetry and new style poetry. In contrast, this study suggests that the evolving characteristics of the Chinese language have profound influence on classical Chinese poetry. It emphasizes the importance of studying classical Chinese poetry (and other Chinese literatures) in the evolutionary context of the Chinese language.

In general, the result of phylogenetic analysis indicates that the pattern of character usage of a particular type of poetry may be useful in estimating the timing (i.e., dynastic period) of a particular type of poetry, i.e., the period from which the poetry was developed. To investigate the power of phylogenetic analysis in estimating the timing of a unknown type of poetry, I collected a data set of 316 poems from the book *Three Hundred Tang Poems* (Sun 1763) and combined this data set with the data set used in the previous phylogenetic analysis. A Neighbor-Joining tree was reconstructed from the merged data set, in which the lineage of the poems from the book *Three Hundred Tang Poems* is grouped with *Tang shi*, indicating that the phylogenetic analysis of character frequencies can successfully identify the style (i.e., timing) of poetry.

## 3. Discussions

In this study, an extensive investigation has been conducted on character usage of classical Chinese poetry, using various statistical analyses to find and compare the patterns of character usage in the poems of nine period genres. The result indicates that each of nine period genres has unique patterns of character usage. The analysis on the general pattern of character preference implies a decreasing trend in use of the Chinese characters that rarely occur in modern Chinese literature along the timeline of dynastic types of classical Chinese poetry. In addition, the similarity scores of character frequencies are employed to measure the distance between different types of classical Chinese poetry. The phylogenetic analysis based on the distance measure suggests that the evolutionary linkages of different types of classical Chinese poetry are congruent with their chronological order. This result indicates that character frequencies contain useful phylogenetic information to infer the evolutionary linkages among various types of classical Chinese poetry.

The statistical analyses in this study are performed with the frequencies of Chinese characters. Other measures or the combination of multiple measures could be adopted to conduct the same analyses in order to improve the accuracy of phylogenetic tree estimation. The data set can also be extended to include modern Chinese poetry. The analysis of the extended data set can help to understand the critical textual changes of Chinese poetry in its transition from classical style poetry to modern style poetry.

The present study focuses on the evolution of classical Chinese poetry. However, the statistical analyses conducted in this study can be generalized to analyze the data sets of general Chinese literature. Such analyses will provide insights about the evolution of general Chinese literatures. This study is an example of using statistical analysis to understand the evolutionary linkages among Chinese literatures.

Table 1: Classification and characteristics of classical Chinese poetry

| Period | Style | Characteristics |
|---|---|---|
| Western Zhou - the Spring and Autumn period | *shi jing* | Four-character lines, from northern china |
| *Zhan guo* | *chu ci* | Irregular line lengths, from southern china |
| Han, Wei, Jin | *Yue fu* (ancient style poetry) | Five-character or seven-character lines, unregulated |
| Tang - Qing | new style poetry | Regulated poetry |

Table 2: The number of classical poems and Chinese characters sampled from each period genre.

| Period | Source | Number of poems | Number of characters |
|---|---|---|---|
| Western Zhou - Spring and Autumn period | shi jing | 305 | 30,384 |
| Zhan guo | chu ci | 15 | 26,529 |
| Han dynasty | Book of Han and Records of the grand historian | 675 | 27,668 |
| Jin dynasty | Book of Jin | 1821 | 77,715 |
| Tang dynasty | Quan Tang Shi | 5000 | 306,848 |
| Song dynasty | Quan Song Shi | 5000 | 260,622 |
| Yuan dynasty | Yuan Shi Bie Yang Ji and Quan Yuan Ci | 8651 | 575,968 |
| Ming dynasty | Lie Chao Shi Ji | 14602 | 1,277,619 |
| Qing dynasty | Qing Shi Hui | 27801 | 2,055,017 |

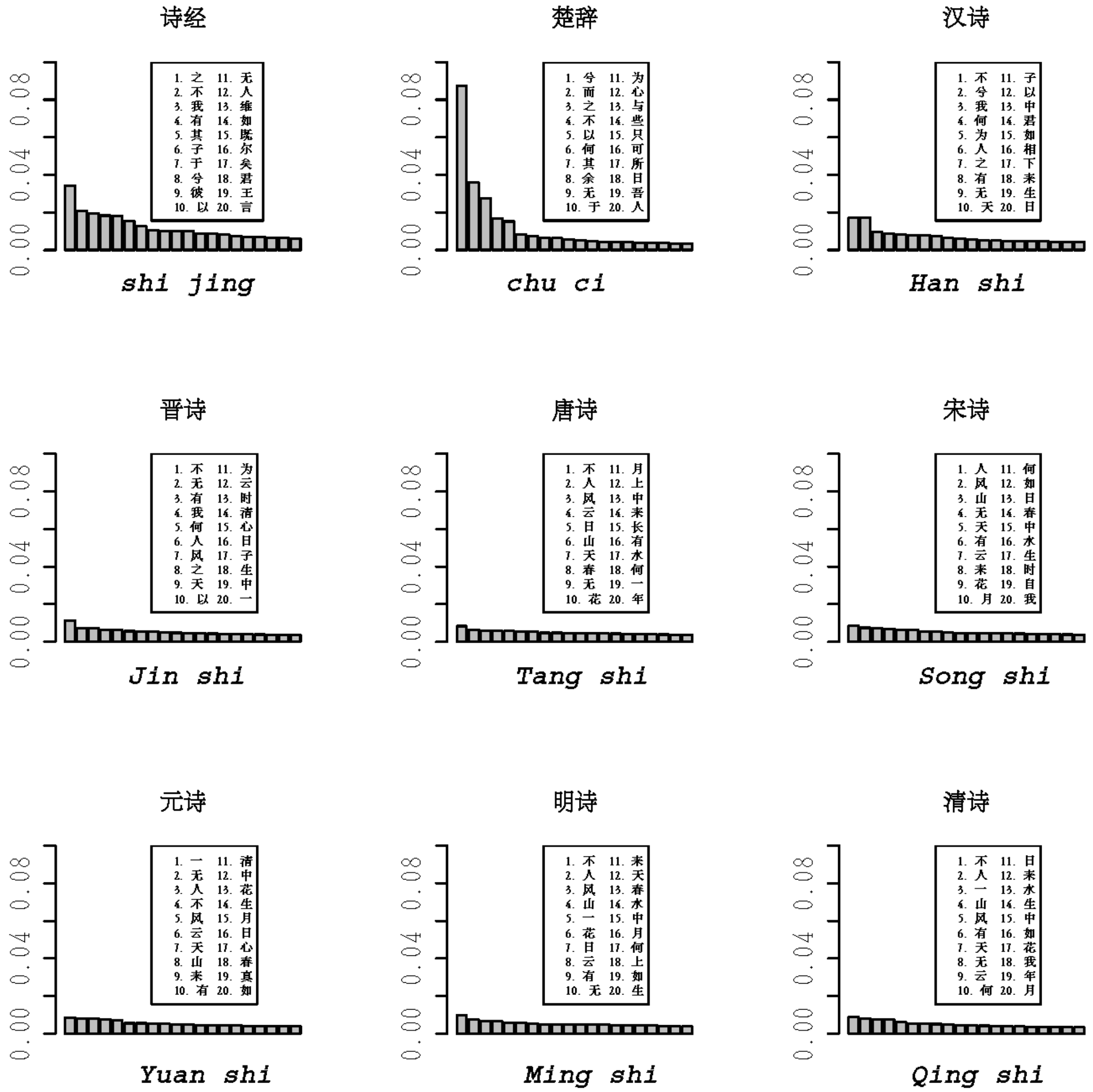

Figure 1：The distribution of Chinese characters for each of nine period gentres, including *shi jing*, *chu ci*, *Han shi*, *Jin shi*,*Yuan shi*, *Ming shi*, and *Qing shi*. The heights of bars represent relative frequencies (i.e., probabilities) of Chinese characters. In addition, the most frequently used characters are listed in descending order.

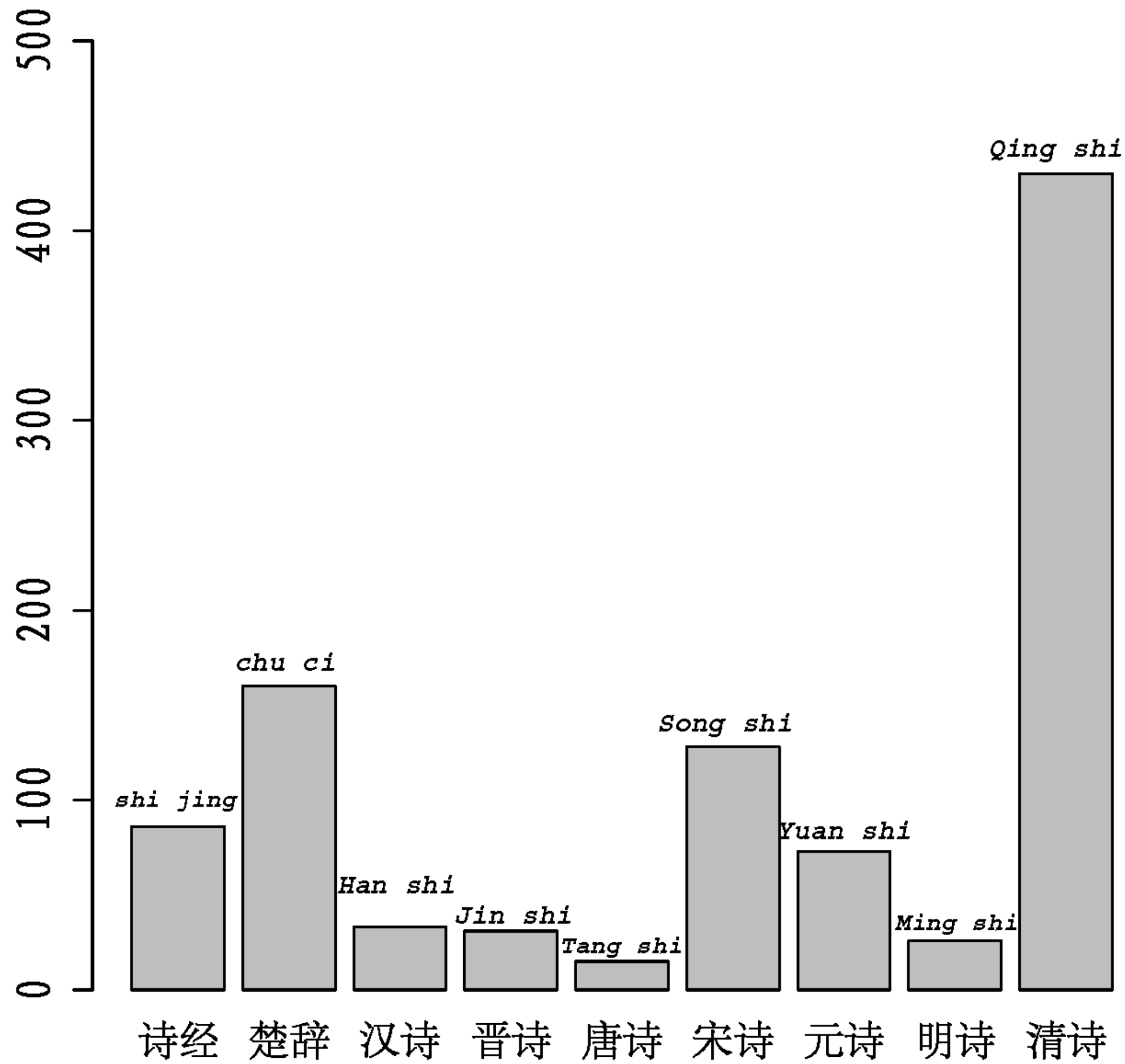

Figure 2: The number of characteristic characters in the poems of nine period genres. The height of a bar represents the number of characteristic characters.

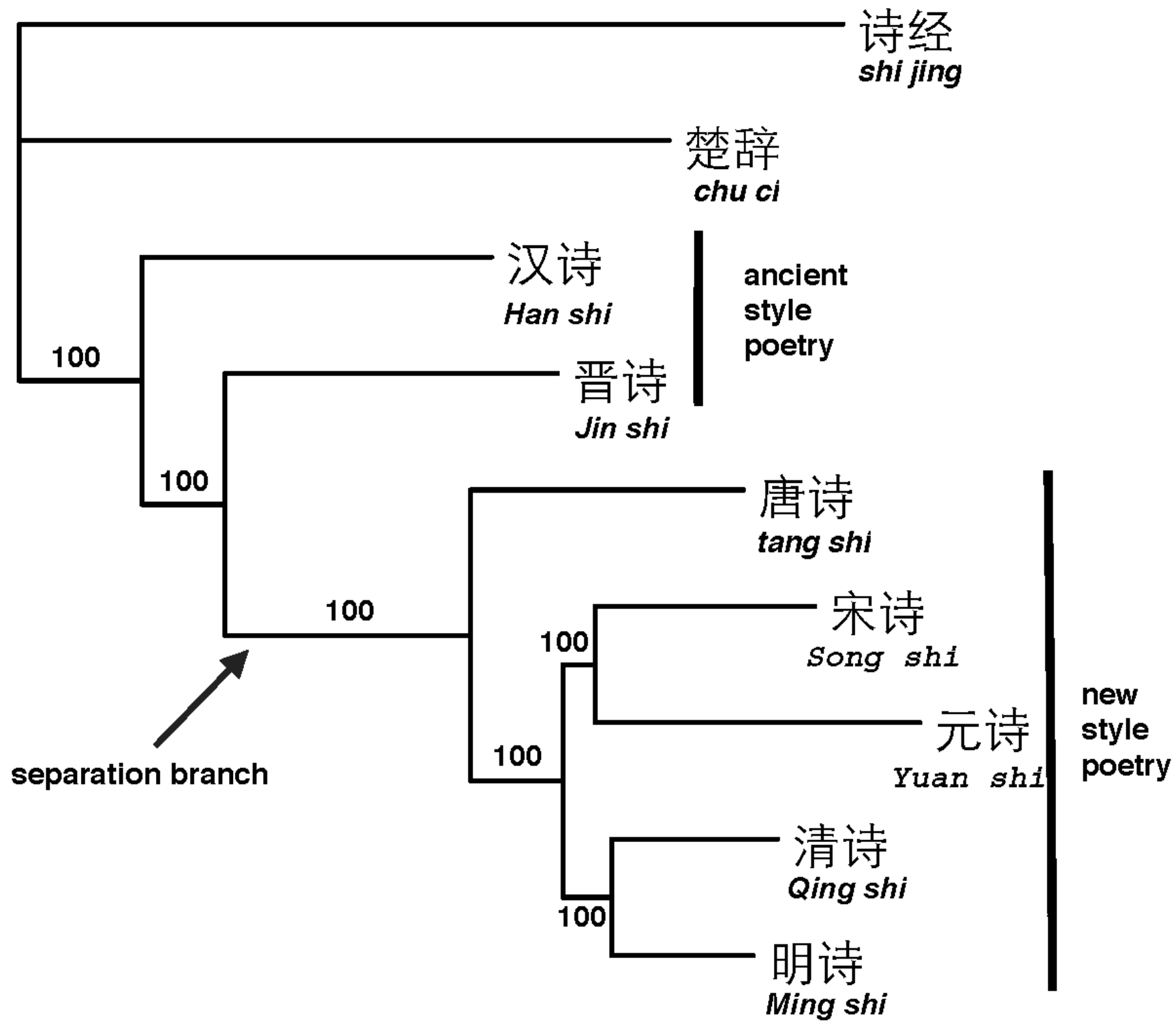

Figure 3: The phylogenetic tree of nine period genres. The branch lengths represent the distances between the poems of nine period genres. The topology of the phylogenetic tree represents the evolutionary linkages among nine period genres. The tree includes two major groups (ancient style poetry and new style poetry) separated by a long branch. The values on the internal branches are bootstrap support values.

## REFERENCES

ANDREEV, S. 2003. Estimation of similarity between poetic texts and their translations by means of discriminant analysis. Journal of Quantitative Linguistics **2**: 159-176.

BIRRELL, A. 1993. Popular songs and ballads of Han China. Honolulu, University of Hawaii Press.

CHENG, C. C. 1991. Quantifying affinity among Chinese dialects. Journal of Chinese Linguistics Monograph Series **3**: 78-112.

DEO, A. S. 2007. The metrical organization of Classical Sanskrit verse. Journal of Linguistics **43**(1): 63-104.

DOBSON, W. A. C. H. 1964. Linguistic Evidence and the Dating of the Book of Songs. Toung Pao 通报 **51**(4): 322-334.

DUNNETT, C. W. 1955. A multiple comparisons procedure for comparing several treatments with a control. Journal of the American Statistical Association **50**(272): 1096–1121.

EFRON, B. 1971. Bootstrap methods: another look at the jackknife. The Annals of Statistics **7**(1): 1-26.

FANG, M. 方旻. 2014. Zhong Hua Shi Ci 中华诗词库 (Classical Chinese Poetry database). Fang Min Gong Zuo Shi 方旻工作室 (Fangmin's Lab).

FREEDMAN, D. A. and W. S.-Y. Wang 1996. Language polygenesis: a probabilistic model. Anthropological science **104**(2): 131-138.

GRAHAM, A. C. 1977. Poems of the Late Tang. New York, The New York Review of Books.

GRIEDER, J. B. 1980. Hu Shih and the Chinese renaissance: liberalism in the Chinese revolution, 1917-1937. Cambridge, Harvard University Press.

HAWKES, D. 2011. The Songs of the South: An Ancient Chinese Anthology of Poems. London, Penguin Books.

HINTON, D. 2008. Classical Chinese Poetry: An Anthology. New York, Farrar, Straus, and Giroux.

HOWELL, D. C. 2007. Statistical Methods for Psychology. Belmont, Thomson Higher Education.

IRMA, S. 2007. Different translations of one original text in a quantitative & qualitative perspective. Exact Method in the Study of Language and Text. P. Grzybek and R. Köhler. Berlin/New York, Gruyter**:** 611-622.

LEE, J. and T.-s. Wong. 2012. Glimpses of Ancient China from Classical Chinese Poems. Proceedings of COLING: 621-632.

LI, D. 李铎. 2005. Quan Song Shi database 全宋诗库 (The poetry of Song dynasty), Peking Univeristy Press.

LIU, H. 刘海涛 and W. Huang 黄伟. 2012. Jiliangxue de xianzhuang, lilun yu fangfa 计量语言学的现状，理论与方法 (Quantitative Linguistics：State of the Art, Theories and Methods). 浙江大学学报 (人文社会科学版) Journal of Zhejiang University (Humanities and Social Science) **43**(2): 178-192.

LIU, J. Y. 1966. The art of Chinese poetry. London, Routledge & Kegan.

MAIR, V. H. and T.-L. Mei. 1991. The Sanskrit Origins of Recent Style Prosody. Harvard Journal of Asiatic Studies **51**: 243-255.

OLGA, L. 2013. Traditional Chinese Poetry. Oxford Bibliographies in Chinese Studies **DOI: 10.1093/OBO/9780199920082-0005**.

PENG, G., J. W. Minett and W. S.-Y. Wang. 2008. The networks of syllables and characters in Chinese. Journal of Quantitative Linguistics **15**: 243-255.

POPESCU, I., R. Cech and G. Altmann. 2011. Vocabulary richness in Slovak poetry. Glottometrics **22**: 62-72.

POPESCU, I., R. Cech and G. Altmann. 2012. Some Geometric Properties of Slovak Poetry. Journal of Quantitative Linguistics **2**: 121-131.

POPESCU, I., R. Cech and G. Altmann. 2013. Descriptivity in Slovak Lyrics. Glottotheory **4**(1): 92-104.

POPESCU, I., L. Mihaiela, D. Tatar and G. Altmann. 2015. Quantitative Analysis of Poetic Texts. Berlin/Boston, de Gruyter.

QIAN, Q. 钱谦益. 2007. Lie Chao Shi Ji 列朝诗集 (Poetry of Ming Dynasty), Zhong Hua Shu Ju 中华书局 (Publisher of China).

QIAO, S. and W. S.-Y. Wang. 1998. Evaluating phylogenetic trees by matrix decomposition. Anthropological science **106**(1): 1-22.

ROXAS-VILLANUEVA, R. M., M. K. Nambatac and G. Tapang. 2012. Characterizing English poetic style using complex networks. International Journal of Modern Physics C **23**.

SAITOU, N. and M. Nei. 1987. The neighbor-joining method: a new method for reconstructing phylogenetic trees. Molecular Biology and Evolution **4**(4): 406-425.

SUN, Z. 孙洙. 1763. Tangshi sanbai shou 唐诗三百首 (Three hundred Tang poems), Zhong Hua Shu Ju 中华书局 (Publisher of China).

VOIGT, R. and D. Jurafsky. 2013. Tradition and Modernity in 20th Century Chinese Poetry. NAACL Second Workshop on Computational Linguistics for Literature.

WANG, W. S.-Y. 1998a. Language and the Evolution of Modern Humans. The Origins and Past of Modern Humans. K. Omoto and P. V. Tobias, World Scientific**:** 247-262.

______. 1998b. Three windows on the past. The Bronze Age and Early Iron Age Peoples of Eastern Central Asia 508-534. (V.H.Mair, ed.), University of Pennsylvania Museum Publications.

______. 2013. The three scales on the evolution of languages. Chinese scientists **1**: 16-20.

WATSON, B. 1971. CHINESE LYRICISM: Shih Poetry from the Second to the Twelfth Century. New York, Columbia University Press.

WATSON, B. 1984. The Columbia Book of Chinese Poetry: From Early Times to the Thirteenth Century. New York, Columbia University Press.

XU, S. 徐世昌. 1929. Qing shi hui 清诗汇 (The collection of Qing poems).

YEH, M. M.-h. 1991. Modern Chinese Poetry Theory and Practice Since 1917. New Haven, Yale University Press.

YIN, C. 曹寅, D. Peng 彭定求, S. Shen 沈三曾, Z. Yang 杨中讷, S. Wang 汪士鋐, D. Wang 汪繹, M. Yu 俞梅, S. Xu 徐树本, D. Che 车鼎晋, C. Pan 潘从律 and S. Zha 查嗣瑮. 1706. Quan Tang Shi 全唐诗 (Collection of Tang poems), Zhong Hua Shu Ju 中华书局 (Publisher of China).

YIP, W.-l. 1976. Chinese poetry: major modes and genres, Univ of California Press.

YIP, W.-l. 1997. Chinese Poetry: An Anthology of Major Modes and Genres. Durham and London, Duke University Press.

ZHANG, Jing-xing 张景星, Yao, Pei-qian 姚培谦, Wang, Yong-qi 王永祺. 1745. Yuan shi bie cai ji 元诗别裁集 (The collection of Yuan poems)

ZHONG, C. 2010. The modern interpretation of classical poetry - rhythm, sentence, Poetry. Chinese Literature and Philosophy of Communication **20**(1): 1-45.

# 浅析中国古代诗歌文字使用的历史变化

中国古代不同类型诗歌之间存在着明显的演变关系。这一点已经被很多关于中国古代文学的研究所共识。然而对中国古代诗歌演变的量化研究却十分有限。这篇文章的主要目的是通过数据分析来量化中国古代诗歌在用字上的演变。具体地讲，运用不同的统计分析手段来比较和发现九种不同时期的古代诗歌（诗经，楚辞，汉诗，晋诗，唐诗，宋诗，元诗，明诗，清诗）在用字上的规律。研究结果表明，不同时期的古代诗歌都有各自独特的用字特点。特定时期的诗歌偏好使用特定的字。不同朝代的诗，按照从诗经到清诗的顺序，逐渐减少了对生僻汉字的使用。另外，研究结果表明不同时期诗歌的演化关系和它们的时间顺序是一致的。进化树分析把古代诗歌分成四个阶段，（诗经，楚辞），（汉诗，晋诗），（唐诗，宋诗，元诗），和（明诗，清诗）。本文使用的统计分析可以应用到其他中国古代文学作品，从而对其他中国文学作品的演变提供量化的证据。